\documentclass[8pt,aps]{revtex4}
\usepackage{amssymb}
\usepackage{latexsym}
\usepackage{epsfig}

\usepackage{dcolumn}
\usepackage{amsmath}
\usepackage{amsfonts}
\usepackage{bm}

\newcommand{\be}{\begin{equation}}
\newcommand{\ee}{\end{equation}}
\newcommand{\bea}{\begin{eqnarray}}
\newcommand{\eea}{\end{eqnarray}}

\newcommand{\p}{\partial}

\newcommand{\la}{\langle}
\newcommand{\ra}{\rangle}
\newcommand{\rd}{\mbox{d}}
\newcommand{\ri}{\mbox{i}}
\newcommand{\re}{\mbox{e}}

\begin{document}
\title{ An integrable model with parafermion zero energy modes}
\author{A. M. Tsvelik\\
Department of Condensed Matter Physics and Materials Science, Brookhaven National Laboratory, Upton, NY 11973-5000, USA}
 \date{\today } \begin{abstract} 
Parafermion zero energy modes are a vital element of fault-tolerant  topological  quantum computation. Although it is believed that such modes form on the border between topological and normal phases, this has been demonstrated only for Z$_2$ (Majorana) and Z$_3$ parafermions.  I consider an integrable model where such demonstration is possible  for Z$_N$ parafermions with any $N$. The procedure is easely generalizable for more complicated symmetry groups. 
\end{abstract}

 %{Insulators }

\maketitle
%\section{General remarks}
 Non-Abelian anyons possess the most exotic statistics known to man. Their permutation  transforms one ground state into another one locally indistinguishable from the first \cite{bais},\cite{goldin}. In conformal field theories this property appears as  non-trivial braiding of the conformal blocks \cite{bzp} (see also \cite{knizhnik}). There are possible applications of non-Abelian statistics  to fault-tolerant  topological  quantum computation \cite{kita03},\cite{freedman1},\cite{freedman2},\cite{nssf08}, but it is also interesting on its own right. 

The simplest anyons  emerge in models of Majorana fermions which material realization has possibly been already achieved \cite{kuwen}. The conceptually simplest and most straightforward generalization of Majorana fermions
is  Z$_N$ parafermions. The former ones have Z$_2$ symmetry and the parafermions have Z$_N$ (N$ >$2) symmetry.  In quantum computation applications one needs to manipulate parafermionic  zero energy modes for which end several schemes have been recently suggested \cite{epl},\cite{shtengel}. In systems with many anyon zero modes they will unavoidably interact so that the degeneracy will be lifted. This is also an interesting subject of research and many lattice models of interacting anyons have been considered (see, for example \cite{ludwig},\cite{gils},\cite{frahm} and references therein).

The  problem number one in this business is how to obtain  anyon zero energy modes. It has been argued, in direct analogy with the Majorana zero modes, that they   emerge on a boundary between ground states with different topological properties (see, for example, \cite{shtengel}). The problem is, however, that  for $N >2$ the parafermions  are interacting objects which makes a consideration of  inhomogenious cases  difficult. So far  the existence of the zero modes  was demonstrated only for N=3 case which can be treated by the Abelian bosonization \cite{shtengel}. 

 Here I suggest a solvable model which contains an inhomogeneity of the required type. The analysis of  the corresponding Bethe ansatz equations supports the idea that a boundary between topologically different states does contain parafermionic zero modes. Although it is not a proof that parafermion zero modes {\it always} emerge on the boundary between ground states of different topology, but this is at least a demonstration that they may emerge there. The derivation is easily generalizable to parafermions from other simple  Lie groups such as, for instance, SU$_k$(N) (see, for example, \cite{mira}). I also derive an effective model describing a finite density of such modes and obtain its exact solution. The latter solutions allows one to estimate the interaction strength  between the parafermions.

The field theoretical definition and properties of massless  Z$_N$ chiral parafermionic fields $\psi,\psi^+$ and $\bar\psi, \bar\psi^+$ can be extracted from the SU$_N$(2) Kac-Moody algebra. The corresponding current operators can be defined in terms of free chiral fermion fields $R,R^+$ and $L,L^+$:
\bea
J^a = \sum_{k=1}^NR^+_{k\alpha}S^a_{\alpha\beta}R_{k\beta}, ~~ \bar J^a = \sum_{k=1}^NL^+_{k\alpha}S^a_{\alpha\beta}L_{k\beta},
\eea
where $S^a$ are spin S=1/2 matrices.
 On the other hand these currents can be written as  \cite{fatzam}
\bea 
&& J^+ = \frac{\sqrt N}{2\pi}\re^{\ri\sqrt{8\pi/N}\varphi}\psi, ~~ J^- = \frac{\sqrt N}{2\pi}\re^{-\ri\sqrt{8\pi/N}\varphi}\psi^+\label{left}\\
&& J^z = \ri\sqrt{N/2\pi}\p_z\varphi\nonumber\\
&& \bar J^+ = \re^{-\ri\sqrt{8\pi/N}\bar\varphi}\bar\psi^+, ~~ \bar J^- = \re^{\ri\sqrt{8\pi/N}\bar\varphi}\bar\psi\label{right}\\
&& \bar J^z = -\ri\sqrt{N/2\pi}\p_{\bar z}\bar\varphi\nonumber,
\eea
where $\psi,\psi^+$ are chiral parafermion fields and $\varphi, \bar\varphi$ are chiral components of the bosonic scalar field $\Phi = \varphi + \bar\varphi$ governed by the Gaussian action 
\bea
S = \frac{1}{2}\int \rd^2x(\p_{\mu}\Phi)^2.
\eea
From (\ref{left},\ref{right}) we deduce the  expressions for the two-point correlators of the parafermion fields 
\bea
\la\la \psi(z)\psi^+(0)\ra\ra \sim z^{-2(1-1/N)}, ~~ \la\la \bar\psi(\bar z)\bar\psi^+(0)\ra\ra \sim {\bar z}^{-2(1-1/N)}.
\eea
which for $N >2$ reveal their nontrivial braiding properties. 

 In a direct analog to the Majorana fermions one can introduce a mass term for the Z$_N$ parafermions. The corresponding action is 
\bea
S = Z_N[\psi,\bar\psi] - \lambda\int \rd^2x[\psi\bar\psi + \psi^+\bar\psi^+], \label{fateev}
\eea
where Z$_N$ term describes the critical part  of the parafermion action. For $N>2$ this is an interacting theory though its properties can be studied since it is integrable\cite{fateev}.
 
In the  N=2 case it is easy  to study a situation where $\lambda$ is coordinate dependent. It is well known that when $\lambda(x)$ changes sign (a kink) the Schr\"odinger  equation has a zero energy  solution where the eigenfunction is localized at the kink (zero energy Majorana bound state).  An important question is whether such bound states exist for N$>$2.  Here I suggest an indirect approach demonstrating existence of the parafermion zero modes. 

 Let us consider the SU$_N$(2) Wess-Zumino-Novikov-Witten model perturbed by the anisotropic current-current interaction. The Hamiltonian density is 
\bea
{\cal H} = \frac{2\pi}{N+2}\left[:J^aJ^a: + :\bar J^a\bar J^a:\right] + g_{\parallel} J^z\bar J^z + g_{\perp}\left[J^+\bar J^- + J^-\bar J^+\right] \label{wznwH}
\eea
The Hamiltonian (\ref{wznwH} describes  the SU(2)-invariant  subsector of the fermionic model with U(1)$\times$U(1)$\times$SU(N) symmetry. This fermionic model was solved by the Bethe ansatz \cite{tsv87}. At $g_{\parallel} >0$ the theory is massive and has solitons and antisolitons.  As is evident from the exact solution, each (anti)soliton carries parafermion  zero mode which supplies it with the non-Abelian statistics. The corresponding S-matrix in the soliton sector is a tensor product of the XXZ S-matrix (the scattering matrix of the sine-Gordon model) and the RSOS (Restricted Solid on Solid) one \cite{fedya}. At sufficiently large $g_{\parallel}$ it also has soliton-antisoliton bound states. 

The Lagrangian density for Hamiltonian (\ref{wznwH}) can be written as 
\bea
{\cal L} = \frac{1}{2}(\p_{\mu}\Phi)^2 + Z_N[\psi,\bar\psi] -\lambda\left(\re^{\ri\beta\Phi}\psi\bar\psi + H.c.\right), \label{boson} 
\eea
where  $\lambda \sim Ng_{\perp}$ and  $\beta$ is related to $g_{\parallel}$ so that at small couplings we have $\beta^2 = (1+ Ng_z/4\pi)^{-1}$. 

 The last term in (\ref{boson}) is similar to the last term in (\ref{fateev}) where  the role of static function $\lambda(x)$ is played by  dynamic field $\exp[\ri\beta\Phi]$. Since this field changes sign on а soliton configuration  one can use model (\ref{boson}) as a substitute for the model of parafermions (\ref{fateev}) with a coordinate dependent mass gap provided one meets certain requirements. First, the solitons must be slow and on average be far from each other. Second, quantum fluctuations of the bosonic exponent should be small which requires small $\beta$. 

 These requirements are met in the following set up. Let us  apply a magnetic field (it is coupled to the bosonic sector) which strength is slightly below the soliton mass threshold. The field breaks the symmetry between the solitons and the antisolitons. The magnitude of the field is slightly below the soliton mass $М$ so that  
\be
T << M-H <<M.\label{limit}
\ee
 At that temperatures we have a  rarified gas of thermally excited slow solitons and no antisolitons. By looking at the thermodynamic Bethe ansatz (TBA) equations we can establish whether solitons carry parafermionic zero modes.

 The corresponding TBA describing the soliton sector of the theory in the limit (\ref{limit}) can be extracted, for example, from \cite{tsv95}. They are a part of a more general system of equations which may  contain also massive solition-antisoliton bound states  (see \cite{tsvS}) which are irrelevant for the present discussion. 
\bea
&& F/L = - TM\int \frac{\rd \theta}{2\pi}\cosh\theta \ln(1 + \re^{\epsilon_N(\theta)/T}), \label{F}\\   
&& \epsilon_j = Ts*\ln(1+\re^{\epsilon_{j-1}/T})(1+\re^{\epsilon_{j+1}/T}) + Ts*\ln(1+ \re^{\epsilon_N/T})\delta_{j,N-1}, ~~ j =1,...N-1,\label{BA2}\\
&& \epsilon_N - K*T\ln(1 + \re^{\epsilon_N/T}) = - M\cosh\theta +H + Ts*\ln(1+ \re^{\epsilon_{N-1}/T}) + O(\re^{- H/T}). \label{BA1}
\eea
where $L$ is the system size and kernel $K$ is 
\bea
K(\omega) = \frac{\sinh[\pi(\xi -1)\omega/2]}{2\cosh(\pi\omega/2) \sinh(\pi\xi\omega/2)}, ~~ \xi = \frac{1}{8\pi/N\beta^2 -1}.
\eea
\[
s*f(x) = \int_{-\infty}^{\infty} \frac{\rd y f(y)}{\pi\cosh(x-y)}
\]
 I am interested in limit (\ref{limit}). In the first approximation one can replace   quasi-energies $\epsilon_j$ (j =1,...N-1) by their  constant asymptotic values for which  the corresponding integral equations (\ref{BA2}) become algebraic. The solution is 
\bea
1 + \re^{\epsilon_j/T} = \Big\{\frac{\sin\Big[\frac{\pi(j+1)}{N+2}\Big]}{\sin\Big(\frac{\pi}{N+2}\Big)}\Big\}^2.
\eea 
Substituting this into (\ref{F}) we obtain the following expression for the free energy: 
\bea
&& F/L = - TQ\int \frac{\rd p}{2\pi}\re^{- (M-H)/T - p^2/2MT} + O(\exp[-2(M-H)/T]),\label{F1}\\
&& Q = 2\cos\Big(\frac{\pi}{N+2}\Big).
\eea
This expression describes   the free energy of an ideal gas of particles of mass $M$ with a chemical potential $H$. The prefactor $Q$ indicates  that  the state of ${\cal N}$ particles with given energy is degenerate so that in the thermodynamic limit the degeneracy is equal to $Q^{\cal N}$. This degeneracy obviously comes from the parafermionic zero modes bound to the solitons. The fact that $Q$ is not integer is a direct indication that the operators describing zero modes attached to different kinks  do not commute with each other.  For $N=2$ we reproduce the known result $D(2)_{\cal N} = 2^{[{\cal N}/2]}$ for the dimensionality of the Clifford algebra representation of ${\cal N}$ gamma matrices. For $N=3$ the obtained dimensionality is the large ${\cal N}$ asymptotic of Fibonacci numbers:
\bea
\phi = 2\cos(\pi/5) = \frac{1+\sqrt 5}{2}, ~~ D(3)_{\cal N} = [{\phi}^{\cal N} - (-\phi)^{-{\cal N}}]/\sqrt 5. 
\eea

 Expression  (\ref{F1}) is the first term in the expansion of the free energy in the soliton density and, as I have said, describes the ideal gas of anyons. One can move further and extract from (\ref{BA2}) the equations for  interacting anyon gas. The interactions lift the ground state degeneracy. 

 At lowest temperatures we invert the matrix kernel in  (\ref{BA2}) to get the equations in the form where the kernel acts on the term which vanish in T=0 limit:
\bea
T\ln(1+\re^{\epsilon_j/T}) - T{\cal A}_{jk}\ln(1+\re^{-\epsilon_k/T}) = {\cal A}_{j,N-1}*s*T\ln(1+\re^{\epsilon_N/T}), \label{BA3}
\eea
where 
\[
{\cal A}_{jk}(\omega) = 2\coth(\pi\omega/2)\frac{\sinh\{\pi[N-\mbox{max}(j,k)]\omega/2\}\sinh\{\pi\mbox{min}(j,k)\omega/2\}}{\sinh(N\pi\omega/2)}.
\]
At temperatures $T << M$ the distribution function in the right hand side (r.h.s.) of (\ref{BA3}) is very sharp and can be replaced by delta function:
\be
{\cal A}_{j,N-1}*s*T\ln(1+\re^{\epsilon_N(\theta)/T})\approx n_{sol}(T){\cal A}_{j,N-1}*s(\theta), \label{BA4} 
\ee
where $n_{sol}$ is the number of solitons. Then Eqs.(\ref{BA3}) with such r.h.s. look like the TBA for the ferromagnetic XXZ model with $n_{sol}$ sites and anisotropy $\gamma = \pi/N$ with an additional restriction forbidding  solutions with rapidities shifted by $\ri\pi/2$. Such restricted equations describe   the critical RSOS models  \cite{reshetikhin},\cite{resh2} with conformal charge $c = 2(N-1)/(N+2)$. The r.h.s. of  Eqs.(\ref{BA4}) is $\sim n_{sol}$ which means that the bandwidth of the excitations of the interacting anyon gas is proportional to the average distance between the solitons $\lambda \sim n_{sol}^{-1}$. This contradicts a naive expectation that this bandwidth is proportional to the overlap of the zero mode wave functions. Such overlap would be exponentially small  in $M\lambda$. 

  In this paper I used the integrable model to describe a state where the mass term of Z$_N$ parafermions alternates its sign. The analysis of the Bethe ansatz equations show that the parafermions create zero energy bound states attached to the domain walls (solitons). When the density domain walls is finite, the modes interact which  lifts the ground state degeneracy. The characteristic bandwidth of the excitations of this interacting anyon gas is proportional to the density of domain walls.   

 I am grateful to H. Frahm for valuable discussions and interest to the work. The work was supported by the US DOE under contract number DE-AC02-98 CH 10886.

%This is what expected from the parafermion zero modes. 

\end{document}